\documentclass[structabstract]{aa}
\usepackage{graphicx}
\usepackage{amssymb,amsmath}
\usepackage[round]{natbib}

\usepackage{color}
\usepackage{bm}
\usepackage{txfonts}
\usepackage[]{hyperref}

\usepackage{aecompl}

\newcommand{\be}{\begin{equation}}
\newcommand{\ee}{\end{equation}}
\newcommand{\ba}{\begin{eqnarray}}
\newcommand{\ea}{\end{eqnarray}}
\def\slashchar#1{\setbox0=\hbox{$#1$}  % set a box for #1
   \dimen0=\wd0     % and get its size
   \setbox1=\hbox{/} \dimen1=\wd1  % get size of /
   \ifdim\dimen0>\dimen1   % #1 is bigger
      \rlap{\hbox to \dimen0{\hfil/\hfil}} % so center / in box
      #1     % and print #1
   \else     % / is bigger
      \rlap{\hbox to \dimen1{\hfil$#1$\hfil}} % so center #1
      /      % and print /
   \fi}

\begin{document}

%\preprint{ZTF-EP-14-13}
%
\title{A new quark-hadron hybrid equation of state for astrophysics}

\subtitle{I. High-mass twin compact stars}

\author{Sanjin~Beni\'c
\inst{1,2}
\thanks{\email{sanjinb@phy.hr}}
\and
David~Blaschke
\inst{3,4}
\thanks{\email{blaschke@ift.uni.wroc.pl}}
\and
David~E.~Alvarez-Castillo
\inst{4,5}
\thanks{\email{alvarez@theor.jinr.ru}}  
\and
Tobias~Fischer
\inst{3}
\thanks{\email{fischer@ift.uni.wroc.pl}}
\and
Stefan~Typel
\inst{6}
\thanks{\email{s.typel@gsi.de}}
}

\institute{Physics Department, Faculty of Science, University of Zagreb, Bijeni\v cka c.~32, Zagreb 10000, Croatia
\and
Department of Physics, The University of Tokyo, 7-3-1 Hongo, Bunkyo-ku, Tokyo 113-0033, Japan
\and    
Institute for Theoretical Physics, University of Wroc\l aw, Pl. M. Borna 9, 50-204 Wroclaw, Poland
\and
Bogoliubov Laboratory for Theoretical Physics, Joint Institute for Nuclear Research, 141980 Dubna, Russia
\and
Instituto de F\'{\i}sica, Universidad Aut\'onoma de San Luis Potos\'{\i}, S.L.P. 78290, M\'exico
\and
GSI Helmholtzzentrum f\"{u}r Schwerionenforschung GmbH, Planckstra\ss{}e 1, 64291 Darmstadt, Germany
}

\date{Received: day month year; accepted: day month year}

% \abstract{}{}{}{}{}
% 5 {} token are mandatory

\abstract
{}
% aims heading (mandatory)
{We present a new microscopic hadron-quark hybrid equation of state model for astrophysical applications, from which compact hybrid star configurations are constructed. These are composed of a quark core and a hadronic shell with a first-order phase transition at their interface. The resulting mass-radius relations are in accordance with the latest astrophysical constraints.}
% methods heading (mandatory)
{The quark matter description is based on a quantum chromodynamics (QCD) motivated chiral approach with higher-order quark interactions in the Dirac scalar and vector coupling channels. For hadronic matter we select a relativistic mean-field equation of state with density-dependent couplings. Since the nucleons are treated in the quasi-particle framework, an excluded volume correction has been included for the nuclear equation of state at suprasaturation density which takes into account the finite size of the nucleons.}
% results heading (mandatory)
{These novel aspects, excluded volume in the hadronic phase and the higher-order repulsive interactions in the quark phase, lead to a strong first-order phase transition with large latent heat, i.e. the energy-density jump at the phase transition, which fulfils a criterion for a disconnected third-family branch of compact stars in the mass-radius relationship. These twin stars appear at high masses ($\sim$ 2 M$_\odot$) that are relevant for current observations of high-mass pulsars.}
% conclusions heading (optional), leave it empty if necessary
{This analysis offers a unique possibility by radius observations of compact stars to probe the QCD phase diagram at zero temperature and large chemical potential and even to support the existence of a critical point in the QCD phase diagram.}


\maketitle

\section{Introduction}

The physics of compact stars is 
an active subject of modern nuclear astrophysics research since it allows the state of matter to be probed at conditions that are currently inaccessible in high-energy collider facilities: extremes of baryon density at low temperature. It provides one of the strongest observational constraints on the zero-temperature equation of state (EoS) by recent high-precision mass measurement of high-mass pulsars by \citet{Demorest:2010bx} and \citet{Antoniadis:2013pzd}. Any scenarios for the existence of exotic matter and a phase transition at high density that tend to soften the EoS may be abandoned unless they provide stable compact star configurations with a mass not less than $2~M_\odot$.
There are still several possibilities for which it is hard or impossible to detect quark matter in compact stars, namely when a) the phase transition occurs at densities that are too high, exceeding the central density of the maximum mass configuration,  b) the transition only occurs  very close to the maximum mass, beyond the limit of masses for observed high-mass pulsars, or when c) the transition is a crossover (or very close to it) so that the hybrid star characteristics is indistinguishable from that of pure neutron stars. The last case has been dubbed the ``masquerade'' problem \citet{Alford:2004pf}. This case seems to be characteristic of the use of modern chiral quark models with vector meson interactions \citet{Bratovic:2012qs} which are very similar in their behaviour to standard nuclear EoS like APR \citep{APR} or DBHF \citep{Fuchs:2006} in the transition region (see e.g. \citet{Klahn:2006}, \citet{Klahn:2013kga}). However, the opposite case is also possible: when the phase transition to quark matter is accompanied by a large enough binding energy release, corresponding to a jump in density and thus compactness of the star, an instability may be triggered that will eventually  result in the emergence of a third family of compact stellar objects, in addition to white dwarfs and neutron stars. The existence of such a branch of supercompact stellar objects that is {\it \emph{disconnected}} from the neutron star sequence has long been speculated in different contexts related to phase transitions in dense matter~\citep[cf.][]{Gerlach:1968zz,Kampfer:1981yr,Schertler:2000xq,Glendenning:2000gh}.
This phenomenon has been studied as a consequence of the appearance of pion and kaon condensates by \citet{Kampfer:1981yr} and \citet{Banik:2001yw}, respectively; hyperons have been studied by \citet{SchaffnerBielich:2002ki}; and quark matter has been studied by \citet{Glendenning:2000gh}, \citet{Schertler:2000xq}, \citet{Fraga:2001xc}, \citet{Banik:2002kc}, \citet{Agrawal:2009ad}, and \citet{Agrawal:2010er}. All the results of these early studies, however, could be ruled out by the recent observation of high-mass pulsars. The question arose whether the twin star phenomenon as an indicator for a first-order phase transition could also concern compact stars with masses as high as $2~M_\odot$. If answered positively, the observation of significantly different radii for high-mass pulsars of the same mass would also allow conclusions  for isospin symmetric matter as probed in heavy-ion collisions.

The ongoing heavy-ion programs at the collider facilities at RHIC (US) and LHC at CERN in Geneva (Switzerland), combined with the success of modern lattice quantum chromodynamics (QCD), did lead to the result that the nature of the QCD transition at vanishing chemical potential and finite temperature is a crossover. The physics of the QCD phase diagram at finite chemical potential and finite temperatures will be the subject of research within the future high-energy facilities at FAIR in Darmstadt (Germany) and NICA in Dubna (Russia). One of their main goals is to find a critical endpoint (CEP) of first-order transitions or indications for a first-order phase transition at high baryon density like signatures for a quark-hadron mixed phase. In general, a phase transition in isospin asymmetric stellar matter is directly related to the corresponding phase transition in symmetric matter, and therefore relevant to the understanding of the QCD phase diagram ~\citep[cf.][]{Fukushima:2013rx,Fukushima:2014pha}. Since increasing the isospin asymmetry would result in lowering the temperature of the CEP to zero \citep{Ohnishi:2011} the detection of first-order phase transition signals in zero temperature asymmetric compact star matter like the mass twin phenomenon would thus prove the existence of at least one CEP in the QCD phase diagram \citet{Alvarez-Castillo:2013cxa,Blaschke:2013ana}. Unfortunately, heavy-ion collision experiments probe matter with only slight isospin asymmetry of about 60~\% neutron excess and hence cannot provide constraints for larger isospin asymmetry or even $\beta$-equilibrium ($>90~\%$ neutron excess) relevant for compact star phenomenology.

The major ingredient of compact star physics is the zero-temperature EoS in $\beta$-equilibrium~\citep[for recent works, cf.][]{Steiner:2012xt,Masuda:2012ed,Orsaria:2012je,Alford:2013aca,Hebeler:2013nza,Inoue:2013nfe,Klahn:2013kga,Fraga:2013qra,Yasutake:2014oxa,Yamamoto:2014jga}. More precisely, hybrid EoS can be decomposed into three parts: (a) low-density nuclear matter, (b) high-density {\it exotic} matter such as hyperons or quarks, (c) the phase transition region between low- and high-density parts. The conditions for the transition depend on details of the underlying microscopic descriptions of matter. For the EoS to yield a third family  and/or the twin phenomenon, the following two conditions should be fulfilled~\citep[for details, see][]{Haensel:2007yy,Read:2008iy,Zdunik:2012dj,Alford:2013aca}:
\begin{itemize}
\item[(1)] The latent heat of the phase transition should fulfil a constraint 
$\Delta \varepsilon > \Delta\varepsilon_{\rm min}$  \citep{Haensel:2007yy,Zdunik:2012dj,Alford:2013aca}, where $\Delta\varepsilon_{\rm min}\sim 0.6 \varepsilon_{\rm crit}$ 
for the schematic hybrid EoS investigated in \citep{Alford:2013aca} and \citep{Alvarez-Castillo:2013cxa} 
with $\varepsilon_{\rm crit}$ being the critical energy density for the onset of the transition.
\\
\item[(2)] The high-density part of the EoS should be sufficiently stiff.
\end{itemize}
The third family of compact objects is attained via an unstable branch, which can be realized by a soft EoS in the transition region, ensured by  condition (1).  Condition (2) is necessary for the core matter to withstand the pressure from the hadronic shell and thus to provide stability for the new, disconnected hybrid star branch.

Confirming the existence of high-mass twins represents an outstanding challenge for observational campaigns to develop precise radius measurements for compact stellar objects~\citep[cf.][]{Mignani:2012vc,NICER,Miller:2013tca}. If detected, the twin phenomenon would be a compelling astrophysical signature of a strong first-order phase transition in the QCD phase diagram at zero temperature and thus strong evidence for the presence of at least one critical end point. By invoking that the high-mass pulsars PSR~J1614-2230 by \citet{Demorest:2010bx} and PSR~J0348+0432 \citet{Antoniadis:2013pzd}, with their precisely measured masses $1.97\pm0.04$~M$_\odot$ and $2.01\pm0.04$~M$_\odot$, respectively, could be such twin stars, we predict in this work that their radii should differ at least by about 1~km (depending on the model details). It remains to be shown whether these values are within the capabilities of future experimental X-ray satellite missions like the Neutron Star Interior Composition Explorer (NICER)~\footnote{http://heasarc.gsfc.nasa.gov/docs/nicer/index.html}, the Nuclear Spectroscopic Telescope Array (NUSTAR)~\footnote{http://www.nasa.gov/mission-pages/nustar/main}, and/or the Square Kilometer Array (SKA)~\footnote{http://www.skatelescope.org}.

In this work we present a microscopically founded example for the class of hybrid EoS that fulfil   criteria (1) and (2) for
the occurrence of a third family of compact stars based on a first-order phase transition from hadronic matter to quark matter.
In our case, the nuclear matter phase is described by a relativistic mean-field (RMF) model with density-dependent meson-nucleon couplings introduced in \citet{Typel:1999yq} using the DD2 parametrization from \citet{Typel:2009sy} with finite-volume modifications. The quark matter phase is given by a Nambu-Jona-Lasinio model (NJL) with higher-order quark interactions,  as introduced in \citet{Benic:2014iaa}. For the phase transition between hadronic and quark matter phases we apply a Maxwell construction. The resulting quark-hadron hybrid EoS allows for massive twin star configurations for which their gravitational masses are in agreement with the present 2~M$_\odot$ constraint set by \citet{Demorest:2010bx} and \citet{Antoniadis:2013pzd}.

The paper is organized as follows. In Sect.~2 we introduce our new quark-hadron hybrid EoS and in Sect.~3 we discuss
its characteristic features  such as excluded volume, mass-radius relations, and twin configurations. The paper closes with the summary in Sect.~4.

\section{Model equation of state for massive twin phenomenon}

For quark matter at high densities we employ the recently proposed NJL-based model of \citet{Benic:2014iaa}. For the low-density region we use the nuclear RMF EoS~\citep{Typel:1999yq} with the well-calibrated DD2 parametrization of \citet{Typel:2009sy}. In order to maximize the latent heat at the phase transition, we correct the standard DD2 EoS by accounting for an excluded volume of the nucleons that results from Pauli blocking due their quark substructure. The latter aspect will be  introduced in the following subsection.

\subsection{Excluded nucleon volume in the hadronic equation of state}
\label{sec:DD2EV}

The composite nature of nucleons can be modelled by the excluded-volume mechanism as discussed\ by e.g. \citep{Rischke:1991ke} in the context of RMF models. Considering nucleons as hard spheres of volume $V_{\rm nuc}$, the available volume $V_{av}$ for the motion of nucleons is only a fraction $\Phi = V_{av}/V$ of the total volume $V$ of the system.  
The available volume fraction can be written as
\begin{equation}
\label{eq:Phi}
\Phi = 1-v\sum_{i=n,p} n_{i}\,\,\,,
\end{equation}
with the nucleon number densities $n_{i}$ and the volume parameter
\begin{equation}
v = \frac{1}{2} \frac{4\pi}{3} \left( 2 r_{\rm nuc}\right)^{3}  = 4 V_{\rm nuc}\,\,\,,
\end{equation}
if we assume identical radii $r_{\rm nuc} = r_{n} = r_{p} $ of neutrons and protons. The total hadronic pressure and energy density are given by the  relations
\ba
p_{\rm tot}(\mu_{n},\mu_{p}) &=& \frac{1}{\Phi} \sum_{i=n,p} p_{i} + p_{\rm mes}\,\,\,, \\
\varepsilon_{\rm tot}(\mu_{n},\mu_{p}) &=& -p_{\rm tot} + \sum_{i=n,p}\mu_i\,n_i\,\,\,,
\ea
with contributions from nucleons and mesons. They depend on the nucleon chemical potentials $\mu_{n}$ and $\mu_{p}$. The nucleonic pressures are given by
\begin{equation}
p_{i} = \frac{1}{4} \left( E_{i} n_{i} - m_{i}^{\ast} n_{i}^{(s)}\right)\,\,\,,
\end{equation}
with the nucleon number densities and scalar densities
\ba
n_{i} &=& \frac{\Phi}{3\pi^{3}} k_{i}^{3}\,\,\,, \\
n_{i}^{(s)} &=& \frac{\Phi m_{i}^{\ast}}{2\pi^{2}} \left[ E_{i} k_{i} - \left( m_{i}^{\ast} \right)^{2}   \ln \frac{k_{i}+E_{i}}{m_{i}^{\ast}}\right]\,\,\,
\ea
that contain the energies
\begin{equation}
E_{i} = \sqrt{k_{i}^{2}+\left( m_{i}^{\ast} \right)^{2}}  = \mu_{i} -V_{i} -\frac{v}{\Phi}\sum_{j=p,n} p_{j} \: ,
\end{equation}
as well as Fermi momenta $k_{i}$ and effective masses $m_{i}^{\ast}=m_{i}-S_{i}$. The vector potentials $V_{i}$, scalar potentials $S_{i}$ and the mesonic contribution $p_{\rm mes}$ to the total pressure have the usual form of RMF models with density-dependent couplings \citep[for more details, see][]{Typel:1999yq}.

\begin{table}[ht]
\caption{Parameters of the DD2-EV model. The definition of the quantities are given in \citet{Typel:2009sy}.}
\label{tab:DD2EV}
\centering
\begin{tabular}{cccc}
\hline \hline
 meson $i$ & $\omega$ & $\sigma$ & $\rho$ \\
\hline
 $\Gamma_{i}(n_{\rm sat})$ & $12.920700$ & $10.826881$ & $2.296878$ \\
 $a_{i}$ & $1.693762$  & $1.357629$ & $2.215411$ \\
 $b_{i}$ & $-0.002358$ & $0.634443$ &  -- \\
 $c_{i}$ & $0.050349$  & $1.005359$ &  -- \\
 $d_{i}$ & $2.573015$  & $0.575809$ &  -- \\
\hline \hline
\end{tabular}
\end{table}

In conventional RMF models the in-medium nucleon-nucleon interaction is modelled by the exchange of ($\sigma$, $\omega$, and $\rho$) mesons  between pointlike nucleons. The excluded volume causes an additional effective repulsion between the nucleons. Hence, the parameters of the RMF model have to be refitted in order to retain the characteristic properties of nuclear matter. The parameters of the nucleon-meson couplings in the DD2 RMF model were determined by fitting to properties of finite nuclei \citep[for details, see][]{Typel:2009sy}. This approach leads to very satisfactory results all over the nuclear chart and gives nuclear matter parameters that are consistent with current experimental constraints.  In Table \ref{tab:DD2EV} the parameters of the new parametrization DD2-EV with excluded-volume effects are given assuming a volume parameter $v = (1/0.35)$~fm$^{3}$. This corresponds to a nucleon radius of $r_{\rm nuc} \approx 0.55 $~fm. See \citet{Typel:2009sy} for the definition of the quantities in the table and their relation to the coupling functions. The saturation density $n_{\rm sat}$ and the particle masses are not changed  compared to the original DD2 parametrization. The DD2-EV parameters were determined such that the binding energy per nucleon $E/A$, the compressibility $K$, the symmetry energy $J$, and the symmetry energy slope parameter $L$ are also identical to that of the DD2 effective interaction. For the hadronic EoS we use the original DD2 parametrization without excluded-volume effects at baryon densities below the saturation density $n_{\rm sat}$ of the model since these densities are well tested in finite-nucleus calculations. At densities above $n_{\rm sat}$ we replace the DD2 model by the DD2-EV parametrization with excluded-volume corrections. The maximum baryon density that can be described by this model is $n_{\rm max} = 1/v = 0.35$~fm$^{-3}$ owing to the choice of the volume parameter $v$. At this density the pressure diverges and the transition to quark matter has to start below $n_{\rm max}$. In stellar matter the usual contributions to the pressure and energy density of the electrons are added to the hadronic part. Requiring charge neutrality, i.e. $n_{e} = n_{p}$ and $\beta$ equilibrium, where $\mu_{n}=\mu_{p}+\mu_{e}$, the pressure and energy density become functions of a single quantity, the baryon chemical potential $\mu_{B} = \mu_{n}$.

The expected transition from hadronic to quark matter at zero temperature with increasing density is still lacking a full microscopic theoretical description from first principles. In particular, reliable models with both hadronic and quark degrees of freedom are not available. Thus, a phase transition construction with separate models for both phases has to be employed. The critical chemical potential, the type of phase transition, and the extension of a possible mixed phase are not known. The aim of our approach for the EoS is to provide a reasonable microscopic model that allows for an extended mixed phase of hadrons and quarks with a large latent heat. This condition governs the choice of the volume parameter $v$ and corresponding small maximum density $n_{\rm max}$. It does not mean that nucleons cannot be found at higher densities. They exist up to a density of 0.404~fm$^{-3}$ in beta equilibrium in the mixed phase. It is not possible to identify the relevant degrees of freedom unambiguously in a dense medium. From a more fundamental point of view, the quark substructure of nucleons has to be considered. It leads to strong repulsion at high densities due to the action of the Pauli exclusion principle on the quark level.

\subsection{NJL model with 8-quark interactions}

In order to describe cold quark matter that is significantly stiffer than the ideal gas, we employ the recently developed generalization of the NJL model by \citet{Benic:2014iaa}, which includes 8-quark interactions in both Dirac scalar and vector channels (NJL8). 
The mean-field thermodynamic potential of the two-flavor NJL8 model is given as 
\ba
\Omega &=& U -2 N_c\sum_{f=u,d} \Bigg[\frac{1}{2\pi^2}\int_0^\Lambda d p\, p^2 E_f
-\frac{1}{48\pi^2}\Bigg\{(2\tilde{\mu}_f^3-5M_f^2\tilde{\mu}_f)
\nonumber\\
&&\times\sqrt{\tilde{\mu}_f^2-M_f^2}+3M_f^4\ln\Bigg(\frac{\sqrt{\tilde{\mu}_f^2-M_f^2}+\tilde{\mu}_f}{M_f}\Bigg)\Bigg\}
\Bigg]-\Omega_0~,
\label{eq:pot}
\ea
with
\ba
U &=& 2\frac{g_{20}}{\Lambda^2}(\phi_u^2+\phi_d^2) + 12 \frac{g_{40}}{\Lambda^8}(\phi_u^2+\phi_d^2)^2 -2\frac{g_{02}}{\Lambda^2}(\omega_u^2 + \omega_d^2)
\nonumber
\\
&& - 12 \frac{g_{04}}{\Lambda^8}(\omega_u^2+\omega_d^2)^2\,\,\,,
\ea
and energy-momentum relation, $E_f = \sqrt{{p}^2+M_f^2}$, with
\ba
M_u &= &m+4\frac{g_{20}}{\Lambda^2}\phi_u+ 16\frac{g_{40}}{\Lambda^8}\phi_u^3 + 16\frac{g_{40}}{\Lambda^8}\phi_u \phi_d^2\,\,\,,
\label{eq:cmass}
\\
\tilde{\mu}_u &=& \mu_u-4\frac{g_{02}}{\Lambda^2}\omega_u- 16\frac{g_{04}}{\Lambda^8}\omega_u^3 - 16\frac{g_{04}}{\Lambda^8}\omega_u \omega_d^2\,\,\,.
\label{eq:tildm}
\ea
Expressions for $M_d$ and $\tilde{\mu}_d$ are obtained by cyclic permutation of indices in \eqref{eq:cmass} and \eqref{eq:tildm}, respectively. The model parameters are the 4-quark scalar and vector couplings $g_{20}$ and $g_{02}$, the 8-quark scalar and
vector couplings $g_{40}$ and $g_{04}$, as well as the current quark mass $m$ and the momentum cutoff $\Lambda$, which is placed on the divergent vacuum energy. 
The constant $\Omega_0$ ensures zero pressure in the vacuum.

The model is solved by means of finding the extremum value of the thermodynamic potential \eqref{eq:pot} with respect to the mean-fields ($X=\phi_u, \, \phi_d,\, \omega_u,\, \omega_d$), i.e.
\ba
\frac{\partial \Omega}{\partial X} = 0\,\,\,,
\ea
and the pressure is obtained from the relation $p=-\Omega$.

In this work we use the parameter set of \citet{Kashiwa:2006rc}, $g_{20} = 2.104$, $g_{40} = 3.069$, $m=5.5$ MeV, and $\Lambda=631.5$ MeV. Furthermore, the vector channel strengths are quantified by the ratios
\ba
\eta_2 = \frac{g_{02}}{g_{20}} \, , \qquad
\eta_4 = \frac{g_{04}}{g_{40}}~.
\ea
Here, we will concentrate on the parameter space where $\eta_2$ is small and use $\eta_4$ to control the stiffness of the EoS. We note  that small $\eta_2$ ensures an early onset of quark matter, i.e. it refers to low densities for the onset of quark matter (depending on the stiffness of the nuclear EoS at corresponding densities).

Within this approach we can calculate the partial pressures $p_f$ and densities $n_f = \partial p_f/\partial\mu_f$ for $f=(u,d)$. In neutron stars, neutrino-less $\beta$-equilibrium is typically fulfilled, i.e. the corresponding equilibrium weak process in nuclear matter is the nuclear $\beta$-decay: $n\leftrightarrows p+e^- + \bar{\nu}_e$. In quark matter, it is replaced by $d\leftrightarrows u + e^- +\bar{\nu}_e$, and hence here the following relation holds between the contributing chemical potentials, $\mu_d = \mu_u + \mu_e$ (neutrino escapes from the star, so its chemical potential is set to zero). Moreover, the following condition,
\ba
\sum_{i=u,d,e}Z_i\,n_i = \frac{2n_u}{3} - \frac{n_d}{3} -n_e = 0\,\,\,,
\ea
ensures local charge neutrality. The total pressure in the quark phase is then given by the sum of the partial pressures, $p = p_u + p_d + p_e$, with electron pressure $p_e$. The latter is calculated based on the relativistic and degenerate Fermi gas. Moreover, the baryon chemical potential and the baryon density in the quark phase ($Q$) and hadronic phase ($H$) are obtained as 
\ba
\mu_B^Q &=& \mu_u + 2\mu_d \,\,\,,\qquad \mu_B^H = \mu_n\,\,\,, \\
n_B^Q &=& \frac{\partial p}{\partial \mu_B^Q}  = \frac{n_u + n_d}{3}\,\,\,,\qquad n_B^H = n_p + n_n\,\,\,,
\ea
with neutron and proton chemical potentials ($\mu_n,\mu_p$) and densities ($n_n,n_p$).
When no confusion arises, indices $Q$ and $H$ will be omitted for simplicity.

\begin{figure}[!ht]
\includegraphics[width=0.49\textwidth,angle=0]{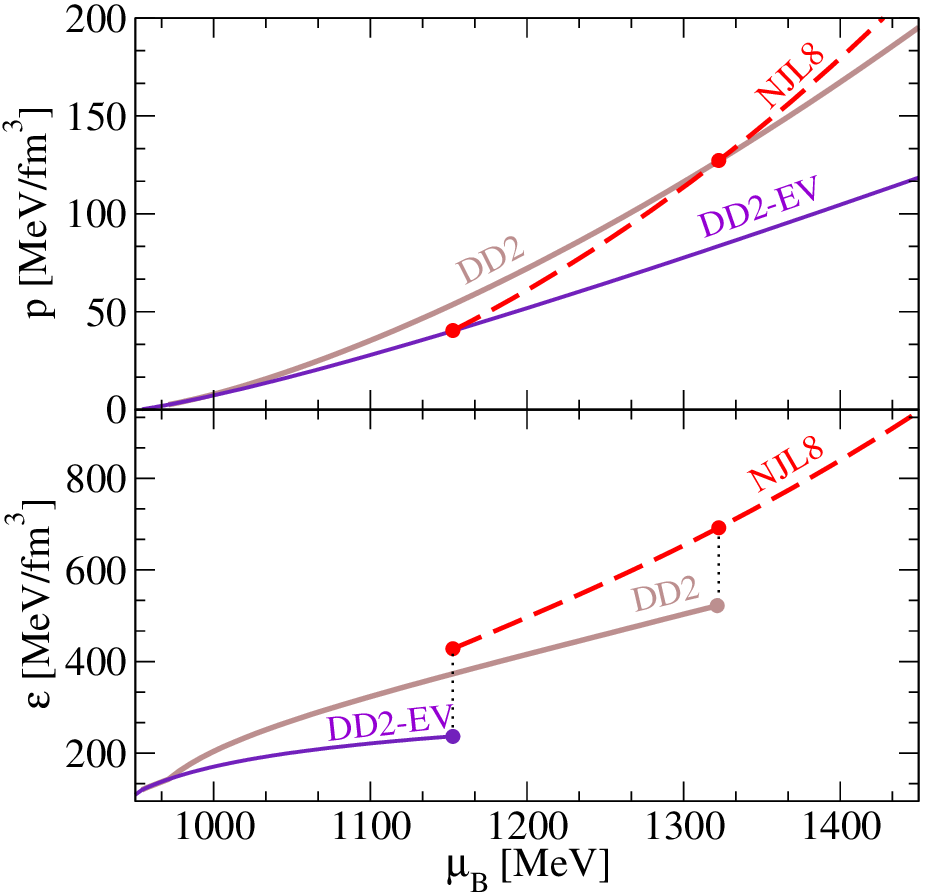}
\caption{\label{fig:eos2}
(Colour online) Upper panel: Maxwell construction in the $P-\mu_B$ plane for the DD2 EoS (brown solid line) and the NJL8 EoS (red dashed line) for $\eta_2 = 0.08$, $\eta_4=0.0$, as well as the DD2-EV EoS (violet solid line) and the same NJL8 EoS parametrization. 
Lower panel: the same construction as in the upper panel for the corresponding $\varepsilon-\mu_B$ plane.}
\end{figure}

For the construction of the phase transition, we apply Maxwell's condition in the pressure-chemical potential plane, i.e. pressures in quark and hadronic phases must be equal $p_H(\mu_B^H)=p_Q(\mu_B^Q)$ at coexistence $\mu_B^H = \mu_B^Q$, in order to ensure thermodynamic consistency. This approach is tantamount to assuming a large surface tension at the hadron-quark interface. The critical baryon chemical potential is obtained by matching the pressures from the hadronic (DD2-EV) and quark (NJL8) EOSs. With this setup, a first-order phase transition is obtained by construction with a significant jump in baryon density and energy density as illustrated in Fig.~\ref{fig:eos2}. It will be  discussed further in   Sect.~3.

\section{Results}

The model parameters used to calculate the hybrid EoS are as follows. We modify the DD2 EoS with the excluded volume mechanism as described in Sect.~\ref{sec:DD2EV} (DD2-EV). 
The high-density part is given by the NJL8 EoS \eqref{eq:pot}, where we use $\eta_2=0.08$ and consider 
$\eta_4$ as a free parameter.

The rationale behind our choice of a low value for $\eta_2$ and the particular value for $v$ is at this stage purely phenomenological. The parameter $\eta_2$ controls both the onset of quark matter and the stiffness of the quark EoS. 
We note that a measure for the stiffness (or softness) of the EoS is the speed of sound $c_s$
defined via
\ba
c_s^2 = \frac{\partial p}{\partial\epsilon} = \frac{\partial \ln \mu_B}{\partial \ln n_B}~.
\ea
When comparing two EoS, the stiffer one has the steeper slope of $p(\varepsilon)$, while the slope of $n(\mu_B)$ is lower.
In the present model,
a larger value for $\eta_2$ would result in more similar quark and hadronic EoS, and hence disfavour the anticipated condition of maximized latent heat at the phase transition. 
Since a small value of $\eta_2$ ensures a low onset of quark matter, the nuclear EoS is insensitive to the detailed behaviour of the $\Phi$ function close to the maximum density $n_{\rm max}$. 
Thus we use the traditional linear dependence \eqref{eq:Phi} on the nucleon densities.

\begin{figure}[!ht]
\includegraphics[width=0.49\textwidth,angle=0]{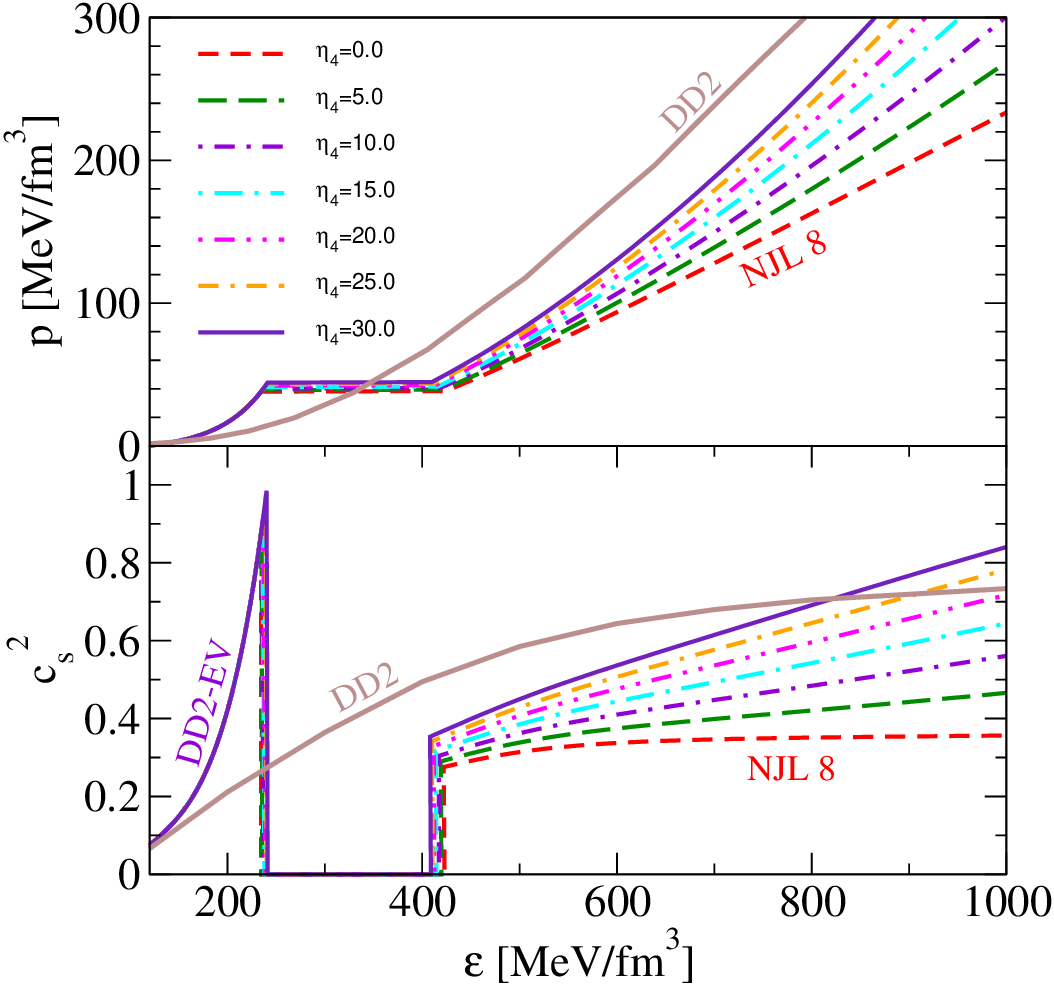}
\caption{\label{fig:eos}
(Colour online) Upper panel: Hybrid EoS built from DD2-EV and NJL8 with phase transition via the Maxwell construction, for the NJL8 parameter $\eta_2=0.08$ and varying $\eta_4$ from 0.0 -- 30. 
Lower panel: squared speed of sound for these hybrid EoS.
For comparison, in both panels the hadronic EoS DD2 is shown (solid brown line). }
\end{figure}

\subsection{Hybrid equation of state}

The new quark-hadron hybrid EoS, based on DD2-EV and NJL8, is shown in the upper panel of Fig.~\ref{fig:eos} illustrating the pressure-energy density plane, for fixed $\eta_2=0.08$ and varying vector-coupling parameters $\eta_4$. 
We note that with increasing vector coupling parameters, $\eta_2$ and $\eta_4$, the sound speed rises (see lower panel of Fig.~\ref{fig:eos}). 
We also note  that the different mass-dimension of the 4-quark and 8-quark vector operators separates the region in density in which the respective operator influences the speed of sound. 
In particular, while $\eta_2$ controls the stiffness of the EoS in the low-density region, $\eta_4$ stiffens the EoS in the high-density region. 
This can be seen from Fig.~\ref{fig:eos} where $\eta_2=0.08$ and $\eta_4$ is varied from $\eta_4 = 0.0$  (red dashed line)
%, 5.0 (dashed, red line), 10.0 (dot-double-dashed green line) and 20.0 (full, black line)
to $\eta_4 = 30.0$ (violet solid line). 
In addition to the selected values shown in Fig.~\ref{fig:eos}, we explored the total parameter range 
$\eta_4=0.0-30.0$ in steps of $\Delta\eta_4=1.0$. 
Above the maximum value of $\eta_4=30.0$ the transition from DD2-EV to NJL8 violates the requirement of causality (Fig.~\ref{fig:eos}), which takes place at $\epsilon \simeq 240$~MeV~fm$^{-3}$. 
We have checked that in all our cases the causality limit is reached only at energy densities beyond which the mass-radius sequences turn unstable.

Exploring the available parameter spaces in both hadronic and quark matter phases, we have found the maximized latent heat in the combination of two aspects: 
(a) taking into account finite-sized effects of the nucleons using the excluded volume and 
(b) applying small values of $\eta_2$ for the NJL8 quark-matter model.  
This is illustrated in Fig.~\ref{fig:eos2} where we compare the phase transition constructions from DD2  and DD2-EV  to NJL8  with $\eta_2=0.08$ and $\eta_4=0.0$. 
The upper panel of Fig.~\ref{fig:eos2} shows pressure vs. chemical potential, from which it becomes clear that our excluded volume approach reduced the critical chemical potential for the onset of quark matter. 
Furthermore, it also increases the differences between the slopes of the pressure curves for hadronic and quark EoS at the phase transition. 
The excluded volume results in an increased latent heat, $\Delta\varepsilon$, which is shown in the bottom panel of Fig.~\ref{fig:eos2}. 
For the NJL8 parameters explored here  ($\eta_2=0.08$, $\eta_4=0.0$), we find $\Delta \varepsilon \simeq 0.34\varepsilon_\mathrm{crit}$
for the transition with DD2 and $\Delta \varepsilon \simeq 0.81\varepsilon_\mathrm{crit}$ for the transition with DD2-EV.

With the given choice of nuclear matter parameters, the excluded volume correction generates a stiff nuclear EoS at suprasaturation densities, close to the limit of causality, i.e. $c_H^2 \simeq 1$. Furthermore, the choice of small $\eta_4$ ensures a soft quark matter EoS at the phase transition densities, i.e. $c_Q^2 \gtrsim 1/3$. 
The resulting maximized jump in energy density at the phase transition from DD2-EV to NJL8 is illustrated in Fig.~\ref{fig:eos} for the parameter range $\eta_4=0.0-30.0$, for which we obtain $\Delta\varepsilon \simeq (0.81-0.70)\varepsilon_{\rm crit}$.

Our approach for the construction of a quark-hadron phase transition with large latent heat extends beyond the phenomenological model of \citet{Zdunik:2012dj} and \citet{Alford:2013aca}, known as ZHAHP. 
In their approach, the latent heat $\Delta \varepsilon$ is a free parameter and the quark EoS is defined by a constant speed of sound $c_Q^2$. 
Nevertheless, in providing as one of the major requirements for the existence of the third family the rule of thumb that the latent heat is around $\Delta\varepsilon \simeq 0.6$, the ZHAHP approach proves to be extremely practical\citep[see e.g.][]{Alvarez-Castillo:2013cxa}. 
However, it is unphysical to treat $c_Q^2$ and $\Delta\varepsilon$ as mutually independent parameters. 
Within a microscopic description for the EoS both quantities are always correlated, for example,  the relative stiffness of the EoS between hadronic and quark phases defines the latent heat
\ba
\Delta \varepsilon = \mu_B^{\rm crit}\left(n_Q - n_H \right)
=(\mu_B^{\rm crit})^2\left(c_Q^2\frac{\partial n_B^Q}{\partial\mu_B}\bigg|_{\mu_B^{\rm crit}}-c_H^2 \frac{\partial n_B^H}{\partial\mu_B}\bigg|_{\mu_B^{\rm crit}}\right)~. 
\ea
In the above formula all the quantities are evaluated at the 
critical chemical potential of the transition $\mu_B=\mu_B^{\rm crit}$.

\subsection{Mass-radius relationship}

\begin{figure}[!ht]
\includegraphics[width=0.49\textwidth,angle=0]{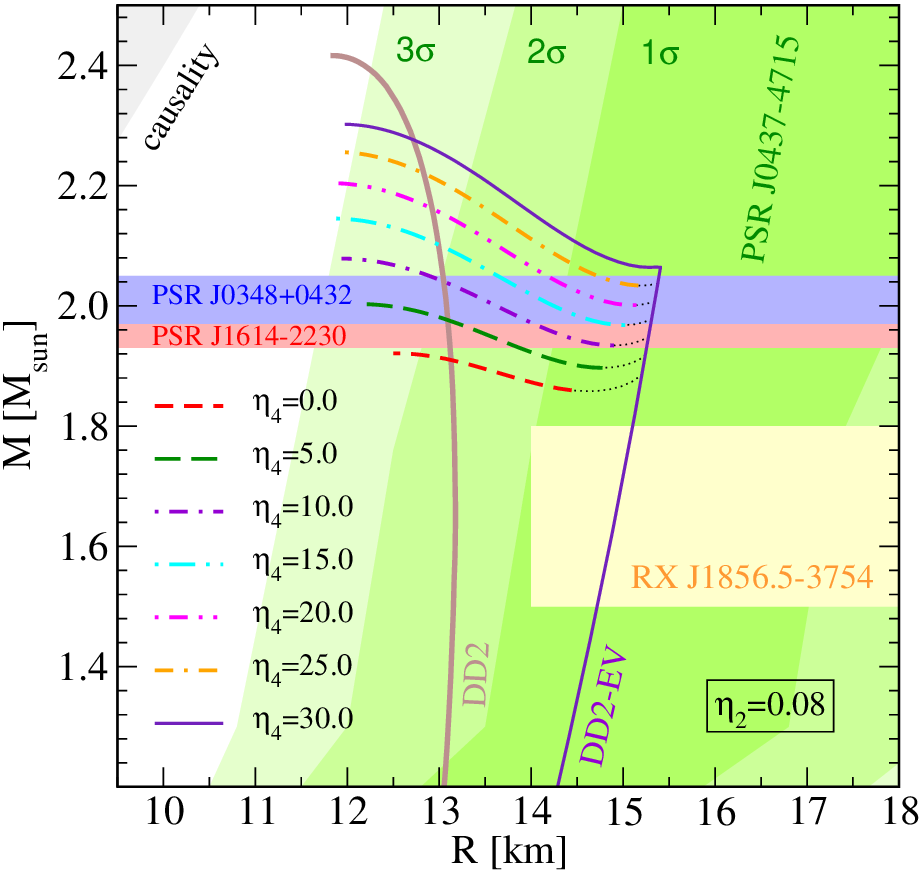}
\caption{(Colour online) Mass-radius relations for our hybrid DD2-EV NJL8 EoSs for constant 
$\eta_2=0.08$ and varying $\eta_4=0.0-30.0$. 
Horizontal colour bands mark the current 2~M$_\odot$ constraints from high-precision mass measurement of high-mass pulsars, PSR~J1614--2230 (red) by \citet{Demorest:2010bx} and PSR~J0348+0432 (blue) by \citet{Antoniadis:2013pzd}. 
For comparison with DD2-EV, the mass-radius curve for the hadronic EoS DD2 is also shown  (solid brown line). 
Furthermore, we show results from the mass-radius analysis of the millisecond pulsar PSR~J0437--4715 by \citet{Bogdanov:2012md}, and constraints from the X-ray spectroscopic study of the thermally emitting isolated neutron star RX~J1856.5--3754 by \citet{Hambaryan:2014via}.}
\label{fig:mr}
\end{figure}

Based on our novel quark-hadron hybrid EoS we calculate the mass-radius relations from solutions of the Tolman-Oppenheimer-Volkoff (TOV) equations. 
For a selection of quark matter parameters, i.e. constant $\eta_2=0.08$ and varying $\eta_4=0.0-30.0$, 
we show the resulting mass-radius curves in Fig.~\ref{fig:mr}. 
Horizontal coloured bands mark the constraints from high-precision mass measurements of the high-mass pulsars PSR~J1614-2230 and PSR~J0348+0432 by \citet{Demorest:2010bx} and \citet{Antoniadis:2013pzd}, respectively. 
In Fig.~\ref{fig:mr}, the green shaded vertical bands mark the results of the mass-radius analysis of the millisecond pulsar PSR~J0437-4715 by \citet{Bogdanov:2012md}, with $1\sigma$, $2\sigma$, and $3\sigma$ confidence level assuming a mass of 1.76~M$_\odot$. 
These data form the basis of a new Bayesian analysis of constraints for hybrid EoS parametrizations
\cite{Alvarez-Castillo:2014xea} which ought to supersede the first study of this kind by \citet{Steiner:2010fz}.
In addition, we show data from the X-ray spin phase-resolved spectroscopic study of the thermally emitting isolated neutron star RX~J1856.5-3754 by \citet{Hambaryan:2014via}, indicating potential compactness constraints. 
The solid brown line in Fig.~\ref{fig:mr} corresponds to the purely hadronic EoS DD2, i.e. without excluded volume corrections, for comparison with DD2-EV.

The   excluded volume approach introduced here results in large neutron star radii, $R\simeq14.75$~km for $M=1.5$~M$_\odot$ in comparison to DD2 ($R\simeq13$~km). 
It can be understood in terms of the significant stiffening of the nuclear EoS above saturation density 
($n_\text{sat}=0.149$~fm$^{-3}$). 
At the phase transition the stellar configuration proceeds from the stable hadronic branch to an unstable branch, marked by dotted lines in Fig.~\ref{fig:mr}. 
The mass-radius coordinates where this happens are defined by the critical chemical potential $\mu_B^\text{crit}$, or density $n_B^\text{crit}$, of the corresponding hybrid EoS. 
We note that for all hybrid EoS explored in this study, the critical density is 
$n_B^\text{crit}\simeq1.5 n_\text{sat}$. 
Specifically, the initially stable hadronic configuration at $\mu_B^\text{crit}$ grows by a tiny amount of mass ($\sim5\times10^{-4}$~M$_\odot$) while the radius remains constant and becomes a still stable hybrid branch. We estimate the size of the resulting quark core to be $\sim80$~cm with significantly increased density. Only after that, the configuration turns to the unstable branch during which the quark core grows. The unstable branch reverts to another stable branch due to the strong repulsive force 8-quark interaction of the NJL8 EoS at high densities.

Our selection of nuclear and quark matter parameters allows not only for high-mass hadronic and quark configurations, in agreement with the 2~M$_\odot$ pulsar data from \citet{Demorest:2010bx} and \citet{Antoniadis:2013pzd}, but also for the consistent transition from the hadronic branch to the quark-hadron hybrid branch. 
Here we identify the latter as the third family of compact stellar objects, with maximum masses in the range $M_{\rm max} = 1.92-2.30$~M$_\odot$, which are above those of the underlying hadronic model DD2-EV. Moreover, we confirm that all hybrid EoS fulfil the condition of causality, i.e. the maximum speed of sound of the hybrid star configurations is in the range $c_Q^2 = 0.34-0.82$. In addition, our results are in agreement with the mass-radius analysis of the millisecond pulsar PSR~J0437--4715 by \citet{Bogdanov:2012md} within $3\sigma$ confidence level (see the green vertical bands in Fig.~\ref{fig:mr}), and  with the compactness study of the isolated neutron star RX~J1856.5--3754 by \citet{Hambaryan:2014via} (see the yellow box in Fig.~\ref{fig:mr}) where a radius of around $14-18$~km at a mass range of $1.5-1.8$M$_\odot$ was found within the $1\sigma$ confidence level \citep[see also][]{Trumper:2011}.

\subsection{Radii difference of the high-mass twins}

The most striking consequence
of a strong first-order phase transition in compact star matter is the possible existence of a third family of compact stars, a branch of stable hybrid star configurations in the mass-radius diagram disconnected from the second family branch of ordinary hadronic stars entailing the twin phenomenon: for a certain range of masses there are pairs of stars (twins) with the same gravitational mass but different internal structure. In order to quantify the unlikeness of the twins as a measure of the pronouncedness of the phase transition we consider the radii difference 
$\delta R=R_{\rm max}-R_{\rm twin}$ between the radius at the maximum mass $M_{\rm max}$ 
on the hadronic branch and that of the corresponding mass twin on the third family branch of hybrid star configurations. 
In Table~\ref{tab:twin}, we list $\delta R$ at fixed $\eta_2=0.08$ for selected values of the dimensionless 8-quark interaction strength $\eta_4$  in the range where it allows for the twin phenomenon (see also Fig.~\ref{fig:mr} for comparison).
\begin{table}[ht]
\caption{Parameters of high-mass twin neutron star configurations for the relevant range of dimensionless 8-quark interaction couplings $\eta_4$ in the vector meson channel at fixed $\eta_2=0.08$. For details see text.}
\label{tab:twin}
\centering
\begin{tabular}{cccccc}
\hline \hline
$\eta_4$ & $M_{\rm max}$  & $R_{\rm max}$  & $R_{\rm twin}$ & $\delta R$  & $\Delta\varepsilon$  \\
 & $[$M$_\odot$$]$ &$[$km$]$ & $[$km$]$ &  $[$km$]$ & $[\epsilon_\text{crit}$]$$ \\
\hline
  0.0 & 1.89 & 15.21 & 13.61 & 1.59 & 0.81 \\
  5.0 &1.92  & 15.24 & 14.09 & 1.16 & 0.79 \\
10.0 &  1.95 & 15.28 & 14.27 & 0.91 & 0.77 \\
15.0 &  1.98 & 15.31 & 14.61 & 0.70 & 0.75 \\
20.0 &  2.01 & 15.34 & 14.82 & 0.52 & 0.73 \\
25.0 &  2.04 & 15.38 & 15.01 & 0.36 & 0.72 \\
30.0 &  2.07 & 15.36 & 15.23 & 0.13 & 0.70 \\
\hline \hline
\end{tabular}
\end{table}

The largest radii difference we obtain for $\eta_4=0.0$, however, is for $M$ below the current maximum mass constraint of \citet{Demorest:2010bx} and \citet{Antoniadis:2013pzd}. 
In agreement with these latter constraints are the parametrizations $\eta_4=5.0-30.0$, with $\delta R=1.16-0.13$~km. 
The reduced radii difference for increasing $\eta_4$ can be understood not only from the stiffening of the quark matter EoS at high densities, but also from the reduced latent heat $\Delta\varepsilon$, i.e. the reduced jump in energy density going from the hadronic EoS to the hybrid EoS (see Fig.~\ref{fig:eos}), also listed in Table~\ref{tab:twin}. 
From the required condition $\Delta\varepsilon>0.6\varepsilon_\text{crit}$, it becomes clear from Table~\ref{tab:twin} that twin configurations are only obtained for $\eta_4=0.0-30.0$. 
For $\eta_4\gtrsim30.0$ the phase transition to quark matter 
proceeds without developing a disconnected 
third family branch; all configurations on this sequence up to the maximum mass are stable 
(see also Fig.~\ref{fig:mr}).

\section{Conclusions}

Compact stars harbour central densities in excess of nuclear saturation density, conditions that are currently inaccessible in nuclear high-energy experiments. Their study contributes to a key direction of research in nuclear and hadron physics, i.e. the possible transition from a state of matter with nuclear degrees of freedom to a deconfined state with quark and gluon degrees of freedom. Despite the success of lattice QCD at vanishing chemical potential and high temperatures identifying the nature of the transition as crossover, for finite chemical potentials only phenomenological models can be used \citep[cf.][and references therein]{Lattimer:2010uk,Klahn:2013kga,Buballa:2014jta}. Such models, in particular with the phase transition from nuclear to quark matter, have also been very useful  in astrophysical applications, e.g. in simulations of protoneutron star cooling~\citep[cf.][]{Pons:2001,Popov:2006,Blaschke:2013} and simulations of core-collapse supernovae~\citep[cf.][]{Sagert:2009,Fischer:2011,Nakazato:2014}. It is therefore of paramount interest to develop quark-hadron hybrid models from which it is possible to deduce observables that allow us to further constrain the as yet highly uncertain QCD phase diagram, e.g. the possible existence of a critical point. This identification will be possible with the discovery of a first-order phase transition at low temperatures and large chemical potential, conditions which refer to the state of matter at compact star interiors in $\beta$-equilibrium.

In this paper, we took on this challenge and developed a novel quark-hadron hybrid EoS. It is based on the nuclear EoS DD2, which is a relativistic mean-field model with density-dependent couplings. While such models treat nucleons as point-like quasi-particles, here we also take   finite size effects of the nucleons into account via an excluded-volume approach above nuclear saturation density. The excluded volume correction introduced here is an attempt to account for the Pauli blocking at the quark level. In \citet{Blaschke:1988} the authors considered Pauli quenching of nucleons as quark substructure effects on the basis of which the strong isospin-dependence has been evaluated (see also \citet{Toro:2006} for the isospin dependence of the phase transition) and applied to obtain stable massive hybrid star solutions \citep[see also][]{Blaschke:1990}. At the current status our excluded volume approach is still quite basic and will be improved in  upcoming studies. For the quark matter EoS we apply the NJL model formalism, including higher-order repulsive quark interactions. This become dominant in particular at high densities. We note that the current status of research for the vector interactions in quark matter remains unsettled \citep[for details, see e.g.][]{Steinheimer:2014kka,Sugano:2014pxa} and its impact on the possible existence of the CEP remains an open question \citep[see][and references therein]{Bratovic:2012qs,Contrera:2012wj,Hell:2012da}. The quark-hadron phase transition has been constructed applying the Maxwell criterion, which results in a strong first-order phase transition. The excluded volume on the hadronic side in combination with the stiff quark EoS results not only in an early onset of quark matter, but also in a large latent heat at the phase transition.

From our novel hybrid EoS which we provide to the community for different values of the higher-order quark interaction strength, we have constructed the mass-radius relations based on TOV solutions. Our main findings can be summarized as follows:
\begin{list}{\labelitemi}{\leftmargin=0.5cm}
\item[(1)] The excluded volume for the high-density nuclear EoS results in large radii for intermediate-mass neutron stars. \\
\item[(2)] The transition to quark matter results in a {\em first} stable hybrid configuration with a tiny quark core, which then turns to the unstable branch. \\
\item[(3)] The unstable branch reverts to a stable hybrid branch owing to the strong repulsive higher-order quark interactions, which we identity as {\em third family} of compact stars
\end{list}
For all configurations explored in this study, we find that the maximum masses belong to the stable hybrid branch and that all EoS remain causal. Moreover, most of our parameter choices fulfil a variety of current constraints on mass-radius relations, such as large maximum masses around 2~M$_\odot$ \citep[][]{Demorest:2010bx,Antoniadis:2013pzd} and radii in the range of 14--17~km for canonical compact objects of $M\simeq1.7$~M$_\odot$~\citep[][]{Bogdanov:2012md,Hambaryan:2014via}.

From an observational perspective, a particularly interesting consequence of a third family of compact objects is the twin phenomenon, where two stars of the same mass have different radii. In the present paper, we even found high-mass twins with $M\simeq2~$M$_\odot$ with radius differences of the order of about 1~km. It remains to be shown whether future surveys that are devoted to neutron star radii determinations, such as the X-ray satellite missions NICER, SKA, and NUSTAR, will have the required sensitivity of less than 1~km and will be able  to resolve the twin phenomenon. It would, in turn, provide a unique signature of a first-order phase transition to exotic superdense matter in compact star interiors.

We note that in addition to the size of the emitting region, the main issue of the radius analyses using X-ray burst sources such as~\citet{Steiner:2010fz} and \citet{Steiner:2012xt}, is the question of the atmosphere composition. If the assumption of a hydrogen atmosphere made by \citet{Guillot:2014} is not  correct and  a helium atmosphere should instead be assumed, then the extracted radius will increase by at least 2~km, see Fig. 3 of \citet{Servillat:2012}. This would then be compatible with stiff neutron star EoS while the mass-radius relation extracted for the hydrogen atmosphere assumption would practically be  compatible only with strange quark matter EoS models. For further details we refer to the reviews by \citet{Miller:2013tca} and \citet{Trumper:2011}.

The aspects discussed in this paper may have important consequences when taken into account consistently in dynamical simulations of supernova collapse and explosions, binary mergers, and so on, where during the phase transition the gain in gravitational binding energy will be available to the system as heat, which in turn can trigger the local production of neutrinos as a result of the different $\beta$-equilibrium condition obtained. Furthermore, the current work improves on the previous phenomenological studies of \citet{Alford:2013aca} and \citet{Alvarez-Castillo:2013cxa}, where the latent heat and the speed of sound were considered as mutually independent parameters.

\section*{Acknowledgements}
We gratefully acknowledge numerous discussions and collaboration work on the topic addressed in this contribution with our
colleagues, in particular with A.~Ayriyan, G.~A.~Contrera, H.~Grigorian, O.~Kaczmarek, T.~Kl\"ahn, E.~Laermann, R.~{\L}astowiecki, M.~C.~Miller, G.~Poghosyan, S.~B.~Popov, M.~Sokolowski, J.~Tr\"umper, D.~N.~Voskresensky and F.~Weber. This research has been supported by Narodowe Centrum Nauki (NCN) within the ``Maestro'' programme under contract number DEC-2011/02/A/ST2/00306. The visits of S.~B. and D.~E.~A.-C. at the University of Wroclaw were supported by the COST Action MP1304 ``NewCompStar'' within the STSM programme. S.~B. acknowledges partial support by the Croatian Science Foundation under Project No.~8799. D.~B. acknowledges support by the Polish Ministry for Science and Higher Education under grant number 1009/S/IFT/14. D.~E.~A-C. is grateful for support by the Heisenberg-Landau programme for collaboration between JINR Dubna and German Universities and Institutes. T.~F. is supported by the NCN within the ``Sonata'' programme under contract number UMO-2013/11/D/ST2/02645. S.~T. acknowledges support by the Helmholtz Association (HGF) through the Nuclear Astrophysics Virtual Institute (VH-VI-417), and by the Heisenberg-Landau programme.

\end{document}